# Beam-beam effects in the Super Proton-Proton Collider

Li-Jiao Wang,[1,2,3]   Tanaji Sen,[4]   Jing-Yu Tang,[1,2,3,*]

[1]*Institute of High Energy Physics, CAS, Yuquan Road 19B, Beijing 100049, China*

[2]*University of Chinese Academy of Sciences, CAS, Yuquan Road 19A, Beijing 100049, China*

[3] *Spallation Neutron Source Science Center, Dongguan 523803, China*

[4]*Fermi National Accelerator Laboratory, P.O. Box 500, Batavia, Illinois, 60510, USA*

Abstract: The Supper Proton-Proton Collider is a next generation hadron collider which is now being designed. A baseline design aims for a peak luminosity of about $1\times10^{35}$ cm$^{-2}$ s$^{-1}$. The focus of this article is the effect of beam-beam interactions which are expected to strongly influence stability in the beams. We start with a discussion of a scheme to generate the crossing angles at the interaction points while also correcting the dispersion thus created. The optics constraints on the achievable $\beta^*$ were studied. Weak-strong simulations were performed to study single particle dynamics via tune footprints, frequency map analysis and dynamic aperture calculations. The long-range interactions with the smallest separations are shown to determine the dynamic aperture. Empirical scaling laws for the dependence of the dynamic aperture on the transverse separations and on the number of long-range interactions are found. A tune scan show several alternative working points with slightly better dynamic aperture than the baseline choice. Finally an option to significantly increase the dynamic aperture by increasing the crossing angle for different choices of $\beta^*$ was studied.

## I. INTRODUCTION

A future hadron collider to be built in China has been proposed over the past decade [1]. This collider will follow after an electron-positron collider will have been built and operated. In this first stage, the Circular Electron Positron Collier (CEPC) is scheduled to explore Higgs physics. In the next stage, the Super Proton-Proton Collider (SPPC) will be planned as an energy frontier collider and as a discovery machine beyond the LHC. CEPC and SPPC will be constructed in the same tunnel of 100 km in circumference. Figure 1 shows the layout of the SPPC [2]. The double-ring collider consists of two long straight sections (LSS) of 4300 m, another six straight sections of 1250 m and eight arc areas. LSS3 and LSS7 are used for the high luminosity proton-proton collisions. LSS1 and LSS5 are used for beam collimation and extraction, respectively. During the CEPC era, LSS1 and LSS5 will be the electron-positron collision areas. In the SPPC, two beams are independently accelerated in their own beam pipes from two opposite directions. They are transported to a common region, the interaction region (IR), of length 310 m and brought to collision with a crossing angle at each interaction point (IP). The protons will be accelerated using four injectors in a chain: a proton linac (p-Linac), a rapid

---

* Corresponding author.

tangjy@ihep.ac.cn

cycling synchrotron (p-RCS), a medium-stage synchrotron (MSS) and the final stage synchrotron (SS) to reach the injection energy of 2.1 TeV for the collider that has a center of mass collision energy of 75 TeV [2].

In the SPPC, the nominal beam-beam (BB) parameter is 0.0075 at each interaction point. There are 82 long-range (LR) interactions and 1 head-on (HO) interaction in each interaction region for nominal bunches [3]. With a crossing angle of 110 μrad, the first parasitic normalized separation in units of the rms beam size is $12\sigma$ and all the other parasitic normalized separations are above $9\sigma$. A future circular hadron collider (FCC-hh) planned at CERN is another next-generation hadron collider. Table I summarizes the main parameters of the SPPC and FCC-hh [4-6]. Comparing the number and separation of long range interactions (LRI) in the two colliders, we observe some crucial differences in the parameters related to the beam-beam interactions. Compared to the FCC-hh, the head-on beam-beam tune shift is larger while LRIs have a smaller range of separations, but are about a third fewer in number in the SPPC. A priori it is not obvious which configuration will lead to a better beam-beam performance so detailed studies are necessary.

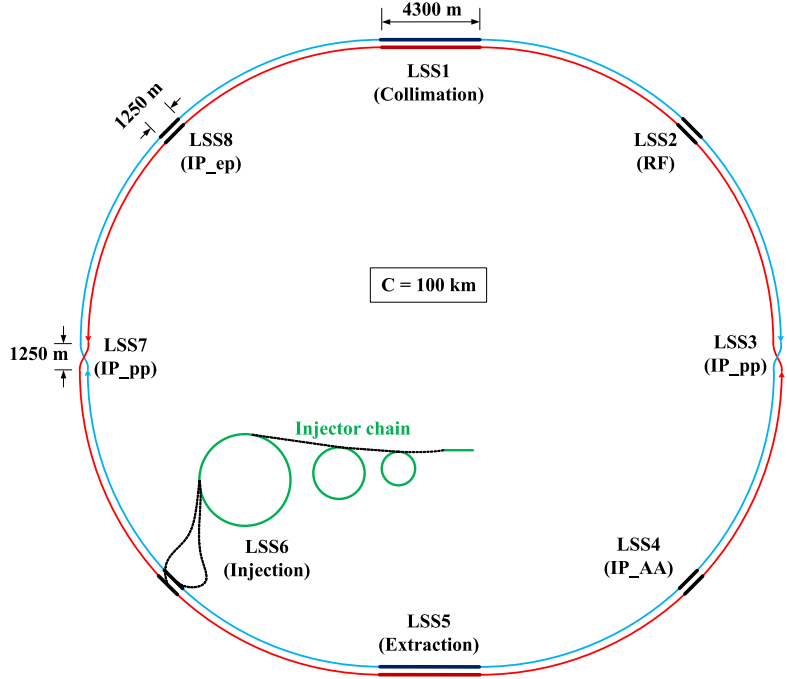

FIG. 1. Layout of the SPPC.

With a bunch population of $1.50 \times 10^{11}$ for 10080 bunches and small beta functions ($\beta^*$) of 0.75 m at the collision points, the peak luminosity can reach $1.20 \times 10^{35}$ cm$^{-2}$ s$^{-1}$ when we assume crab cavities are used to compensate the luminosity loss due to a crossing angle. In the absence of these cavities, the reduction factor due to a crossing angle is 0.85. The hourglass effect can be neglected because rms bunch length of 75.5 mm is much smaller than $\beta^*$.

In the following Section II, we discuss the IR optics with the focus on generating the crossing angle and simultaneously correcting the dispersion generated in each IR.

We study the beam-beam performance in the baseline optics with tune footprints, frequency map analysis (FMA), dynamic aperture (DA) calculations and explore scaling laws in Section III. This is followed in Section IV with a tune scan to find better operating points. In Section V we discuss the impact of increasing the beam separations on the dynamic aperture and we end with our conclusions in Section VI. Coherent beam-beam effects, which could be important, are not addressed here.

TABLE I. SPPC and FCC-hh main parameters.

| Parameter | SPPC | FCC-hh |
| --- | --- | --- |
| Beam energy at collisions (TeV) | 37.5 | 50 |
| Number of IPs | 2.0 | 2.0 |
| Number of bunches | 10080 | 10400 |
| $\beta^*$ (m) | 0.75 | 0.3 |
| Crossing angle (μrad) | 110 | 200 |
| Intensity ($10^{11}$) | 1.5 | 1.0 |
| Norm. trans. emittance (μm) | 2.4 | 2.2 |
| Bunch spacing (ns) | 25 | 25 |
| Rms bunch length (mm) | 75.5 | 80 |
| Rms momentum spread ($10^{-5}$) | 7.07 | 8.15 |
| Peak luminosity ($10^{35}$ cm$^{-2}$ s$^{-1}$) | 1.20 | 2.9 |
| Beam-beam parameter of each IP | 0.0075 | 0.0054 |
| Length of common area (m) | 310 | 435 |
| Number of LRIs | 164 | 232 |
| separations at 1st LRIs ($\sigma$) | 12 | 17 |
| Range of separations at LRIs ($\sigma$) | 9-19 | 12-29 |

## II. IR OPTICS

The IR lattice is a typical left/right antisymmetric optics with the same $\beta^*$ in the two transverse planes. The free space from the IP to the first triplet quadrupole is 45 m. The following inner triplet is responsible for producing $\beta^*$ of 0.75 m, which will cause the maximum $\beta$ of 18.6 km in the triplet. Then two separation dipoles separate the beams into their own beam pipes. Next, the outer triplet matches beta and dispersion to the dispersion suppressor. Figure 2 shows the beta functions in the IR with a horizontal crossing angle (LSS3). The 1st order chromaticity in the ring is corrected by sextupoles, but the 2nd order chromaticity correction system has not been designed yet and is not considered here.

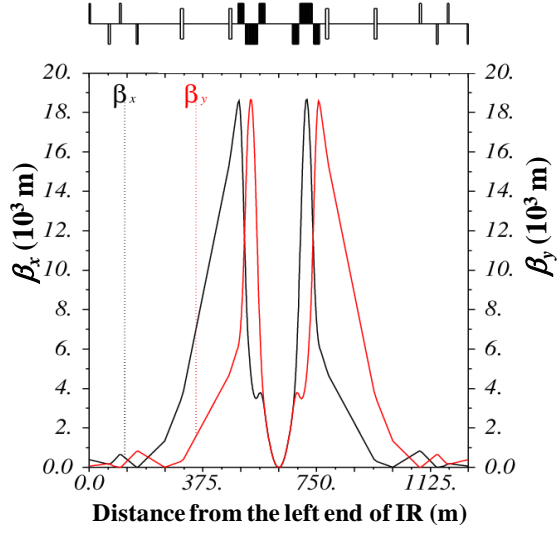

FIG. 2. The beta functions in the IR with a horizontal crossing angle.

The crossing angles at the two interaction points can be created with a set of dipoles [7,8]. In the SPPC, two dipoles on each side of the IP per beam were placed beyond the common area to control the offset and angle at the IP. When the phase advance of a dipole corrector from the IP is $\pi/2$, there is an orbit displacement but no crossing angle at the IP; while if the phase advance is $\pi$, there will be a crossing angle but no orbit displacement at the IP. The present crossing scheme in the two IRs is shown in Fig. 3 with the first orbit corrector placed after the first separation dipole D1. The phase advance of every corrector from the IP is also marked in the plot. In the IR with vertical crossing, the crossing scheme was carried out by vertical orbit correctors. Eight dipole correctors per beam in the two IRs were used to steer the closed orbit.

After adding the crossing angle, the beam orbit is displaced in IR quadrupoles by a maximum of 6.6 mm which leads to unwanted anomalous dispersion at the IP [7,9] which can increase the beam size as shown in the following formula:

$$(\sigma^*_{x,y})^2 = \varepsilon_{x,y} \beta^*_{x,y} + (\sigma_p D^*_{x,y})^2, \quad (1)$$

so that the luminosity will decrease as shown in the following formula:

$$L = \frac{n_b N_b^2 f}{4\pi \sigma^*_x \sigma^*_y} \frac{1}{\sqrt{1+[(\phi \sigma_z)/(2\sigma^*_x)]^2}}. \quad (2)$$

Here, the star superscript implies the value at the IP. $\sigma_{x,y}$, $\varepsilon_{x,y}$, $\beta_{x,y}$ and $D_{x,y}$ are the horizontal and vertical beam size, emittance, beta function and dispersion while $\sigma_p$ is the rms momentum spread. $L$, $N_b$, $n_b$, $f$, $\phi$ and $\sigma_z$ are the luminosity, number of protons per bunch, number of bunches, revolution frequency, crossing angle and rms longitudinal beam size. In addition, with a finite dispersion at the IP, synchro-betatron coupling will be generated and limit the luminosity [10]. The dispersion around the whole ring will increase, especially in the vertical plane, due to the introduction of crossing angles.

We adopted the scheme proposed in Ref. [7] to correct this anomalous dispersion in both planes. Two pairs of quadrupole correctors were placed in the neighboring arc region (where there is horizontal dispersion) on both sides of each IR. In each pair, the two quadrupoles were separated by π in phase had the same strengths and opposite polarities to cancel the changes in beta function from each corrector. For the IR with vertical crossing, a pair of skew quadrupoles was used, because these quadrupoles will transform horizontal dispersion into the vertical plane. Figure 3 shows a sketch of quadrupole correctors in the two IRs. The phase advance of each quadrupole corrector from the IP is marked in Fig. 3.

In Table II, we list the dispersions at both IPs, the maximum horizontal and vertical dispersions in IR3, IR7 and the whole ring. Comparing the 3rd and 4th columns, it shows that the anomalous dispersion has been well compensated by the quadrupole correctors. Figure 4 shows the corrected dispersion in two IRs and its neighboring arc regions. With a corrected dispersion at the IP of 0.001 m, we find using Eq. (2) that the luminosity loss at a momentum spread equal to the rms value is about 0.004% which is negligible.

TABLE II. Dispersion values before and after adding quadrupole correctors at nominal parameters.

| Dispersion ($D_x$, $D_y$) | Without crossing | With crossing but no correctors | With crossing and correctors |
|---|---|---|---|
| ($D_x$, $D_y$) at IP3 (m) | (0, 0) | (0.002, 0) | (0, -0.001) |
| ($D_x$, $D_y$) at IP7 (m) | (0, 0) | (0, -0.002) | (0, 0) |
| Max ($D_x$, $D_y$) in IR3 (m) | (0, 0) | (1.38, 2.43) | (0.03, 0.01) |
| Max ($D_x$, $D_y$) in IR7 (m) | (0, 0) | (2.48, 1.28) | (0.01, 0.04) |
| Max ($D_x$, $D_y$) in ring (m) | (2.8, 0.0) | (2.91, 2.43) | (2.81, 0.29) |

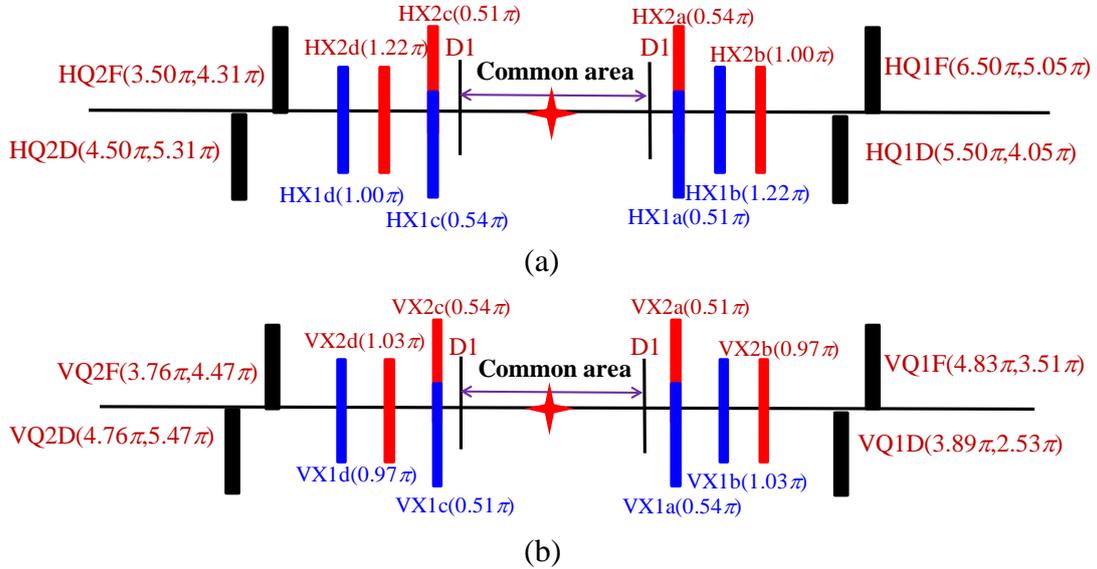

FIG. 3. A sketch of dipole and quadrupole correctors in IR3 with horizontal crossing (a) and in IR7 with vertical crossing (b). D1 is the first separation dipole and the star symbols show the IPs. Blue and red rectangles indicate dipole correctors for beam 1 (blue) and beam 2 (red). The overlapping rectangles show the correctors for both beams in different beam pipes. For dipole correctors, the numbers in brackets are the horizontal (a) or vertical (b) phase advances from the IP. (HX, VX) are the dipole

correctors generating crossing angles in the (horizontal, vertical) plane, respectively. (HQ, VQ) are quadrupole correctors in IR3 and IR7, respectively. (F, D) represents (focusing, defocusing) quadrupoles, respectively. The numbers in brackets are the (horizontal, vertical) phase advances of the quadrupole correctors from the IP.

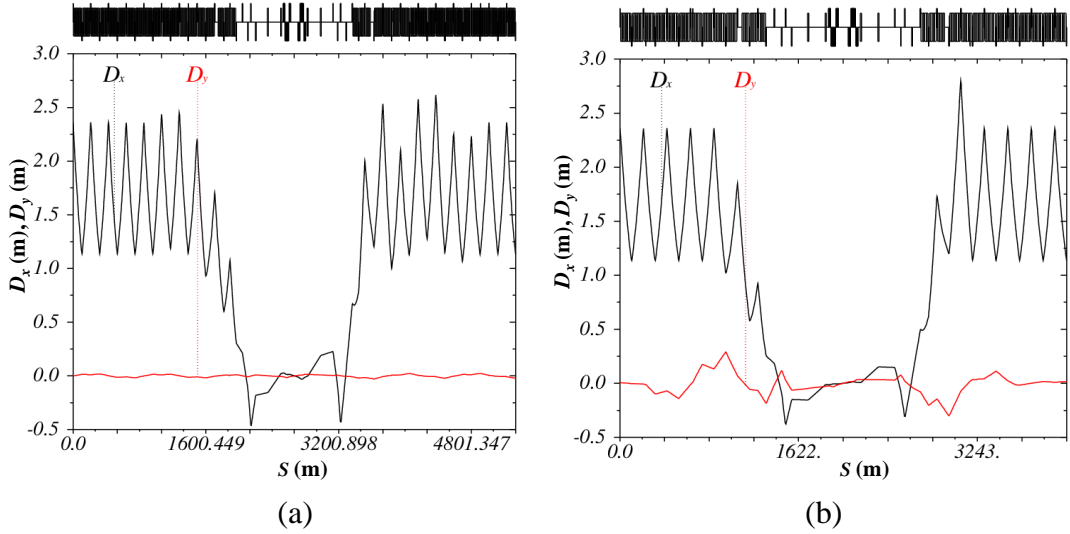

(a)              (b)

FIG. 4. The dispersion in two IRs and its neighboring arc regions with quadrupole correctors: (a) depicts the dispersion in IR3, (b) depicts the dispersion in IR7. The $\beta^*$ is 0.75 m and crossing angle is 110 μrad at both IPs.

The dispersion correction scheme will limit the range of values of the crossing angle and $\beta^*$ achievable while maintaining the required level of correction. We consider here three $\beta^*$ values and different initial parasitic separations. The initial parasitic separation in the drift space before the triplet quadrupoles is normalized by the rms beam size $\sigma$, and can be approximated as shown in the following formula:

$$d = \frac{\theta s}{\sqrt{\varepsilon \beta}} = \frac{\theta s}{\sqrt{\varepsilon(\beta^* + \frac{s^2}{\beta^*})}} \approx \frac{\theta s}{\sqrt{\varepsilon(\frac{s^2}{\beta^*})}} = \frac{\theta \sqrt{\beta^*}}{\sqrt{\varepsilon}}, \quad (3)$$

for $\beta^*$ small enough. In this expression, $\theta$ is the crossing angle, $\varepsilon$ is the geometric emittance, $\beta$ is the beta function at the parasitic interaction and $s$ is the distance of the parasitic interaction location from the IP.

In the first study, the initial separation was kept constant at $12\sigma$, so as $\beta^*$ was varied, the crossing angle scaled as $1/\sqrt{\beta^*}$. Table II shows that the maximum (horizontal, vertical) dispersions $(D_x^{max}, D_y^{max})=(2.8$ m, $0$ m$)$, respectively, without a crossing angle. With the crossing angles, the quadrupole correctors can correct $D_x^{max}$ to about 2.8 m for $\beta^*$=0.5 m, 0.75 m and 1.00 m. The dispersion at the IP is corrected to less than 0.005 m which decreases the luminosity by less than 0.2%. Figure 4 shows the dispersion in the two IRs after correction with the nominal values of $\beta^*$=0.75 m and crossing angle 110 μrad. The correction of the vertical dispersion in IR7 is slightly worse than that of the horizontal dispersion in IR3 because the skew quadrupole correctors are not at the precise phase advances required for the correction. These errors will inevitably be present in the machine, but the quality of the correction does not appear to be significantly affected.

Next we studied the case with increasing the initial separation from $12\sigma$ to $20\sigma$ for the same three values of $\beta^*$ as above. The horizontal dispersion was corrected well in each case but the vertical dispersion $D_y$ was less well corrected. The left plot in Fig. 5 shows the values of $D_y^{max}$ without and with the quadrupole correctors. For both $\beta^*=0.75$ m and 1.00 m, $D_y^{max}$ can be corrected to less than 0.5 m but for $\beta^*=0.5$ m, $D_y^{max}$ exceeds 1 m after correction over the entire range of separations.

The available physical aperture is another important factor in determining the possible range of values of $\beta^*$ and crossing angles. The minimum aperture occurs in the inner triplet at the location of $\beta^{max}$ and decreases linearly with increasing crossing angle. Table III shows some of key parameters when the initial parasitic separation is $20\sigma$. We will assume that the luminosity reduction can be recovered with the use of crab cavities. In the inner triplet, the smallest physical aperture (PA) drops to an unacceptably low value of $9\sigma$ and the beta beating is intolerably large for $\beta^*=0.5$ m. These show that large separations will require $\beta^*>0.5$ m.

TABLE III. Factors limiting the choice of $\beta^*$, the initial separation is $20\sigma$ in all cases.

|  | $\beta^*=0.5$ m | $\beta^*=0.75$ m | $\beta^*=1$ m |
|---|---|---|---|
| Crossing angle ($\mu$rad) | 221 | 184 | 160 |
| Luminosity reduction factor due to crossing angle | 0.55 | 0.69 | 0.79 |
| Max $\beta$ (m) | 28000 | 18660 | 13999 |
| Smallest physical aperture ($\sigma$) | 9 | 14 | 17 |
| $\beta$ beating | 16% | 11% | 9 % |
| Linear chromaticity (before correction) | (-313, -312) | (-259, -258) | (-232, -231) |

We tested the robustness of the dispersion correction scheme by studying its performance in the presence of orbit errors. We assigned alignment errors to the arc quadrupoles that resulted in rms orbit errors of 0.2 and 0.4 mm. This was done for $\beta^*=0.75$ m and 1.00 m over a range of crossing angles. After resetting dipole correctors for the desired crossing angle and zero dispersion at the IP in each case, we find that the maximum dispersion along the ring is basically the same with and without orbit errors. However, due to the non-zero orbit error along the ring, the dipole effects of quadrupoles are excited so that the dispersion correction is affected, the quality of which can be measured by the maximum dispersion in the two IRs. The right plot in Fig. 5 shows the maximum vertical dispersion in the IRs as a function of the initial separation for $\beta^*=0.75$ m. We observe that the maximum vertical dispersion in the IR would lead to a physical aperture loss of less than $0.015\sigma$ at a momentum deviation of $3\sigma_p$. This shows that the dispersion correction would be tolerable for an orbit error of 0.4 mm with $\beta^*$ greater than or equal to 0.75 m. Next we examined the variation in strengths of the correctors compared to the strengths without orbit errors. We find that the maximum variation is 35% which occurs for the HQ2F/HQ2D correctors at a 0.4 mm rms orbit error and the smallest separation of $12\sigma$. This required dynamic range in strengths would appear to be acceptable.

Overall, the dispersion correction scheme appears robust under the set of errors considered here. An rms orbit error of 0.4 mm is acceptable for $\beta^*=0.75$ m and 1.00 m considering both the dispersion correction and the corrector strength variations. However this is a very preliminary study and will have to be reexamined with a more complete set of errors. For example, the magnetic field errors and the misalignment of the IR quadrupoles can increase the induced dispersion at the IP. These high field superconducting magnets are under design and their errors are not known at this time.

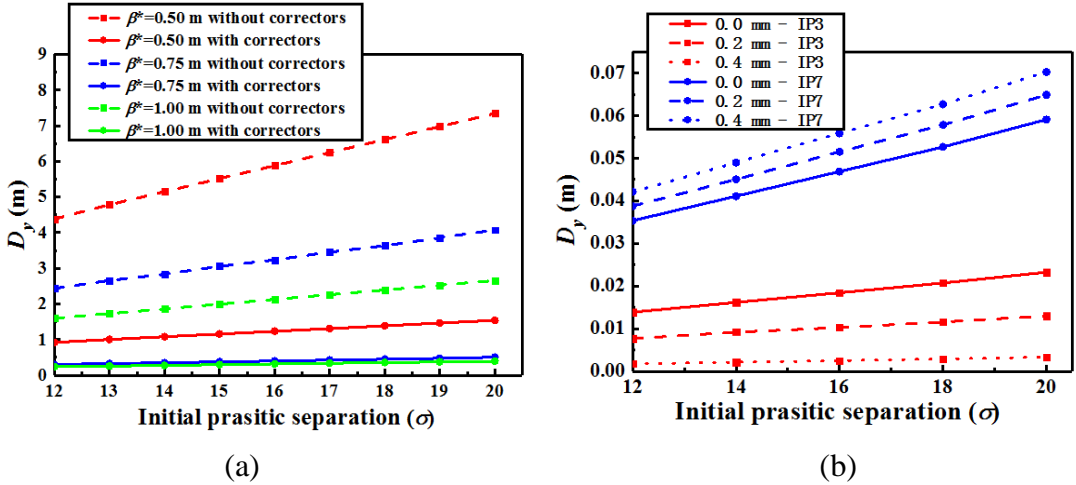

FIG. 5. Left (a): The maximum vertical dispersion $D_y^{max}$ around the ring as a function of the initial parasitic separations before and after adding quadrupole correctors. The dashed (solid) lines show the dispersion before (after) adding the correctors and no orbit errors are included. Right (b): $D_y^{max}$ in the IRs for $\beta^*=0.75$ m vs the initial separations for different orbit errors in the ring. The solid lines represent $D_y^{max}$ without orbit error, the dashed (dot) lines show $D_y^{max}$ with orbit error of 0.2 mm (0.4 mm). Red (Blue) color represents $D_y^{max}$ in IR3 (IR7).

## III. BASELINE SCENARIO PERFORMANCE

Beam-beam interactions have long been known to limit the luminosity in hadron colliders [11-13]. In addition to the head-on interactions, long-range interactions play an important role in limiting the beam lifetime [14-17]. To reduce the effect of the long-range interactions, the SPPC is designed with a short common area where the two beams will share the same beam pipe. Similar to the LHC, the crossing angle is chosen to be in the horizontal plane at one IP and in the vertical plane at the other IP to cancel the tune shifts from the long-range interactions.

Figure 6 shows a plot of the transverse separations at all the BB interactions in LSS3. The transverse separation is normalized by the horizontal transverse beam size. There are 41 long-range interactions (LRIs) on each side of the IP in each IR. The first 12 LRIs which occur before the first quadrupole are at a constant separation of $12\sigma$.

The minimum separation is 9σ-10σ. Finally there are 6 LRIs with constant separation of 17σ on the right side.

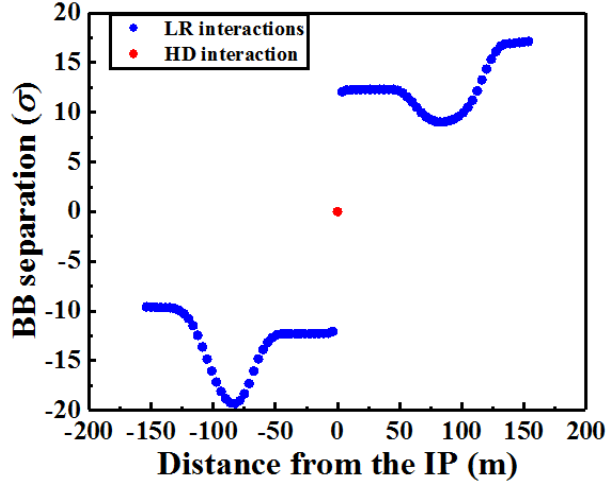

FIG. 6. The BB separations in LSS3 are normalized by their horizontal beam sizes. The blue dots imply the separation of LRIs and the red dot shows the IP.

In addition, bunches in the ring are not always longitudinally spaced by 25 ns. Gaps between different bunch trains are introduced to allow for the rise times of the injection and extraction kickers in the different accelerator stages. In the SPPC, except for the time gap of 3450 ns which is reserved for extraction, the time gaps between different bunch trains are 900 ns. In the SPPC, each beam is distributed in 90 trains of 112 bunches. Any bunch always meets another at each IP, which means all bunches experience 2 HO collisions. At other locations in the common beam pipe, some bunches at the beginning and the end of bunch trains will experience fewer LRIs . Figure 7 shows the number of LRIs experienced by the first two bunch trains in each IR. Only 30 of the 112 bunches in a train experience all 82 LRIs in an IR. The first bunch of the first bunch train misses all the left or right LRIs . Since the gap between different bunch trains (900 ns) is shorter in length than the common area (1025 ns), some of the last bunches in Train 1 will meet some of the bunches at the head in Train 2 of the opposite beam. Thus the last 7 bunches have 47 LRIs and not 41 as for the first bunch in Train 1. Similarly, the last two bunch trains will have the same pattern. In the FCC-hh, the bunch distribution is not uniform because of different gaps in the SPS trains [18]. So the pattern for the number of LRIs experienced by each bunch is more varied.

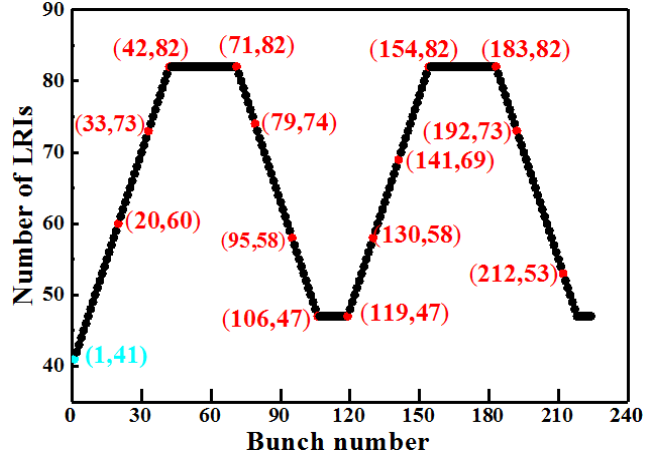

FIG. 7. The number of long-range interactions experienced by bunches in the first two bunch trains. The two numbers ($N_1$, $N_2$) next to the points are the bunch number and the number of parasitics. A Pacman bunch we use later in this paper is marked by cyan color.

In the nominal lattice design, the tunes are (120.31, 117.32), with the fractional parts being the same as in the LHC. The beam-beam parameter is 0.0075 for each interaction point. The weak-strong simulations we now discuss were done using the code BBSIM [19] to study single particle stability with beam-beam interactions. We note that the impact of radiation damping on particle dynamics is not considered in the simulations reported here. We assume that the beam sizes stay constant at the design value corresponding to peak luminosity and particle amplitudes are not damped.

### A. Tune footprints and FMA

The nonlinear tune shift with amplitude due to the beam-beam interactions creates a tune footprint which can be predicted from the theory in Ref. [15]. The theoretical formula to calculate tune shift in the x plane for different amplitude particles is

$$\Delta \nu_x(a_x, a_y, d_x, d_y, r) = \frac{4\pi C}{\varepsilon_x} \int_0^1 \frac{e^{-(p_x+p_y)}}{\upsilon[\upsilon(r^2-1)+1]^{1/2}} \sum_x \sum_y d\upsilon , \qquad (4)$$

where

$$\sum_x = \sum_{k=0}^{\infty} \frac{(\frac{a_x}{d_x})^k}{k!} \Gamma(k+\frac{1}{2})[I_k(s_x)(\frac{2k}{a_x^2}-\upsilon) + I_{k+1}(s_x)\frac{s_x}{a_x^2}] , \qquad (5)$$

$$\sum_y = \sum_{l=0}^{\infty} \frac{(\frac{a_y}{d_y})^l}{l!} \Gamma(l+\frac{1}{2}) I_l(s_y) , \qquad (6)$$

$$C = \frac{N_b r_p}{(2\pi)^3 \gamma_p} , \ r = \frac{\sigma_y}{\sigma_x} , \ p_x = \frac{\upsilon}{2}(a_x^2 + d_x^2) , \ p_y = f\frac{\upsilon}{2}(a_y^2 + d_y^2) , \qquad (7)$$

$$s_x = \upsilon a_x d_x, \; s_y = f \upsilon a_y d_y, \; f = \frac{r^2}{\upsilon(r^2-1)+1}. \tag{8}$$

Here $\upsilon$ is the integration variable in Eq. (4), $a_x$ and $a_y$ are the particle amplitudes normalized by beam transverse size, $d_x$ and $d_y$ are the separations normalized by beam transverse size, $r_p$ is the classical proton radius, and $\gamma_p$ is the relativistic Lorentz factor. $I_n(s_n)$ is the modified Bessel function of the first kind and $\Gamma(n+0.5)$ is the gamma function. There is a similar equation in $\Delta\nu_y$, more details can be found in Ref. [15]. These expressions show that the tune shifts depend on, besides the brightness parameter $N_b/(\gamma_p\varepsilon_x)$, only on the normalized separations, amplitudes and the aspect ratio of the strong beam.

Figure 8 presents the tune footprint calculated from the above theory and the simulations with all of the BB interactions for two different tunes. All sum and difference resonances up to order 4 are also shown. In the simulations, the tune footprint was obtained by tracking particles with amplitude from 0 to $10\sigma_{x,y}$ for 2048 turns, followed by a fast Fourier transform (FFT) to find the tune of every particle. In both plots, the theoretical calculation predicts the main part of tune footprint quite well, especially in the area far from resonances which are not included in the theory. Both plots show differences between the theory and simulations in the area close to the difference coupling resonance. The differences are greater in the left plot at the nominal tune of (120.31, 117.32) due to the presence of the 3rd order resonances nearby. The simulation shows that a few particles are captured by the 3rd order resonance $2\nu_x+\nu_y=1$ which is driven by the long-range interactions. In the right plot at the tunes of (120.17, 117.19) the harmful low order sum resonances are further away and there is no evidence of particle trapping.

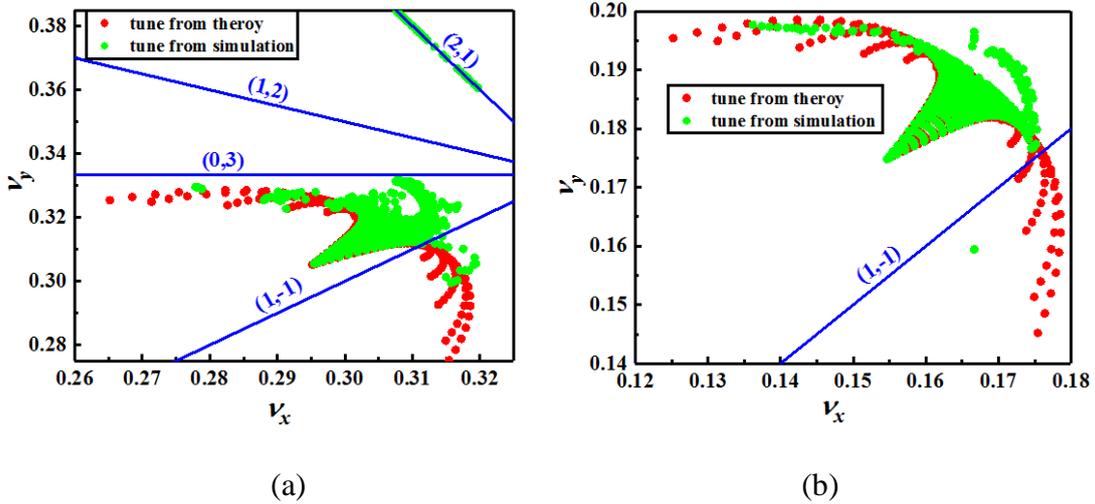

(a)          (b)

FIG. 8. Left (a) footprint for tunes of (0.31, 0.32) and right (b) footprint for tunes of (0.17, 0.19). Red (Green) points are the tunes from the theory (simulations) with all the BB interactions. The lines show all nearby resonances up to order 4. The numbers in brackets (m, n) define resonances $m\nu_x+n\nu_y=p$, where $p$ is an integer.

Figure 9 compares the tune footprints of a nominal and a Pacman bunch. The tune footprint of the Pacman bunch matches that of the nominal bunch except at large amplitudes. This is expected since fewer LRIs will cause weaker effects, while some of the irregular wings are present due to the proximity of the coupling resonance. The tune shift due to the LRIs from IR3 and IR7 compensate each other, the tunes at the bunch core are about the same for both bunches.

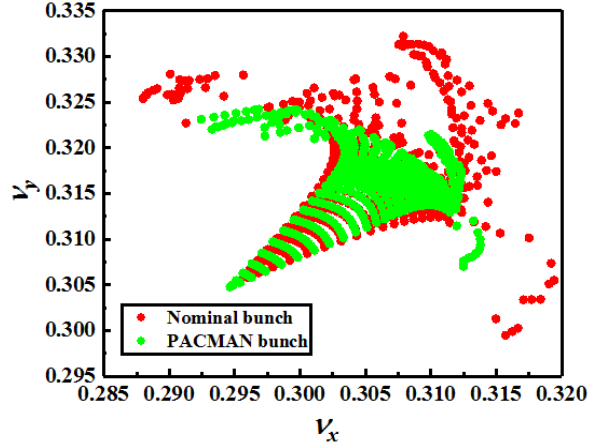

FIG. 9. Red (Green) points are the tune footprint of nominal (Pacman) bunch with all of the BB interactions at the nominal tune.

The FMA method [20,21] was used as an early indicator of particle instability by calculating the diffusion of tunes. Figure 10 shows FMA plots for three cases comparing the effects of head-on and long-range interactions, with sextupole kicks included. Note that the amplitudes extend to $23\sigma$ in the first plot compared with $10\sigma$ in the other two. It clearly shows that the tune diffusion increases much more with the long-range interactions compared with the head-on interactions. The tune variation of particles with amplitude between $6\sigma$ and $10\sigma$ increases by more than four orders of magnitude in the last two plots, which shows that LRIs are the main source of particle instability. Although LRIs mainly impact particles at large amplitude, the small amplitude particles are affected as well.

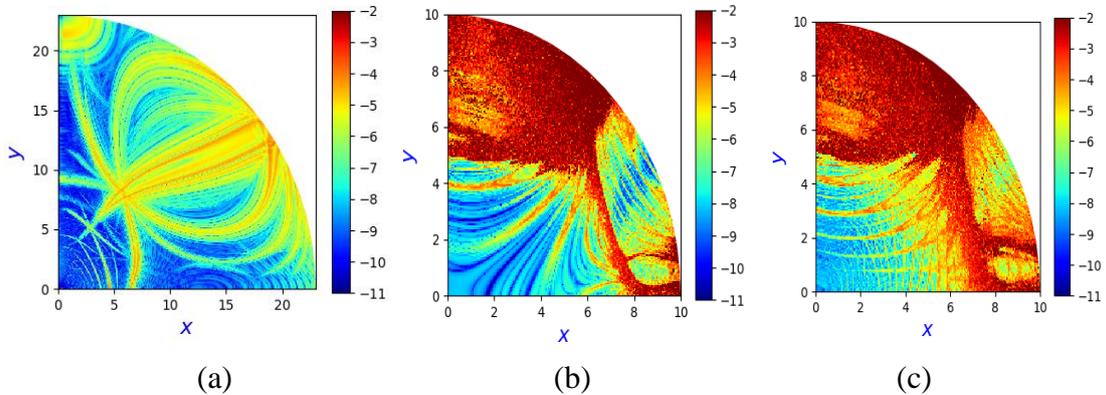

(a)          (b)          (c)

FIG. 10. FMA plots for (a): head-on interactions and sextupole kicks. (b): long-range interactions and sextupole kicks. (c): head-on interactions, long-range interactions and sextupole kicks.

## B. Dynamic aperture

The above short term indicators of particle instability need to be checked with longer term tracking. The dynamic aperture was calculated by tracking particles for $10^6$ turns. Particles were distributed over the phase space $x$, $x'$, $y$, $y'$, $z$ and $dp/p$. The initial transverse distribution was chosen in units of rms transverse beam size ($\sigma$), $x=N\sigma_x Cos\theta$, $y=N\sigma_y Sin\theta$, $x'=0=y'$ and $\theta$ ranged from 0 to $\pi/2$ in steps of 15 degrees. The initial longitudinal distribution in $z$ was a Gaussian truncated at $2\sigma_z$ (rms bunch length) and momentum deviation $dp/p$ was constant at $1\sigma_p$ (rms momentum spread). The chromaticity was corrected to 1.00 here.

We tested four nonlinear settings but not the triplet nonlinearities (those are not yet available). The averaged DA is shown in Table IV. We used the physical aperture as the averaged DA when the simulated averaged DA was larger than the physical aperture. Comparing these four simulation cases for the nominal bunches, the DA is larger than the physical aperture without long-range interactions. The long-range interactions cause the DA down to about $6\sigma$. The addition of the head-on interactions has a relatively small impact on the DA. Figure 11 shows the initial amplitudes of particles (green dots) trapped on the 3rd order resonance $2\nu_x+\nu_y=1$, driven by the LRIs, and most are just outside the dynamic aperture and therefore lost. Therefore, it once again demonstrates that the LRI is the main source of particle loss among the interactions included. The same conclusion that LRIs are more serious compared with head-on interactions can be made for the Pacman bunch.

TABLE IV. Averaged dynamic aperture for 4 different cases. The physical aperture is $23\sigma$.

|  | DA-nominal ($\sigma$) | DA-Pacman ($\sigma$) |
|---|---|---|
| Sextupole | 23 | 23 |
| Sextupole + head-on | 23 | 23 |
| Sextupole + long-range | 6.2 | 9.8 |
| Sextupole + head-on + long-range | 5.5 | 7.8 |

To control the impact of long-range interactions, it's necessary to further study which kind of LRI plays a crucial role among the total 164 LRIs for the nominal bunch. In Fig. 11, we consider the impact of the 48 interactions at $12\sigma$ separation (left plot) and of the 36 interactions at $9\sigma$-$10\sigma$ separations (right plot). The red lines in both plots show that the averaged DA is smaller when only the 36 smallest separation LRIs are included compared to that when only the 48 LRIs are included. Similarly, the cyan lines show that the averaged DA is larger when the smallest separation LRIs are excluded compared to excluding the more numerous constant LRIs. In addition, comparing the cyan and red curves in the right plot show that the DA with only 36 LRIs is comparable to the DA with all the remaining 128 LRIs. These observations imply that the LRIs with the smallest separations are dominant and their impact on the DA is nearly the same as from all the other LRIs.

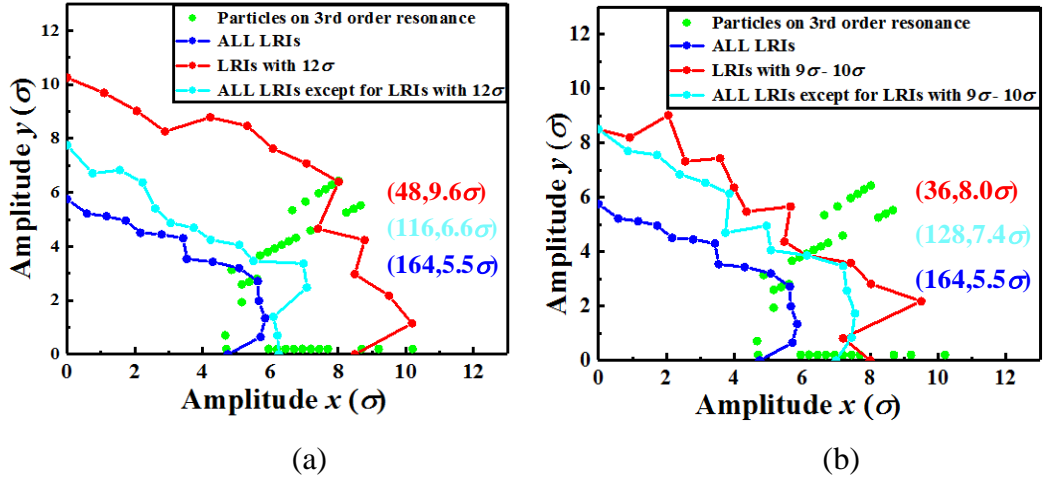

(a)                  (b)

FIG. 11. DA in amplitude space for different numbers of LRIs. The blue lines show that DA with all of the interactions in both plots. The left (a) plot shows the DA when the 48 LRIs at constant separations of $12\sigma$ only are included (red) and when they are excluded (cyan). The right (b) plot shows the same data but for the 36 LRIs at the smallest separations. The green dots show the initial amplitudes of particles that are trapped on the 3rd order resonance $2\nu_x+\nu_y=1$. In parenthesis are (number of LRIs, averaged DA).

It is possible that the DA values obtained with the chosen model are conservative estimates because the radiation damping effects are not included. However, this needs to be verified in future studies.

### C. Scaling laws

The DA could be increased with optics optimization and other changes to machine parameters, for example, increasing bunch spacing and bunch intensity to have fewer LRIs, increasing the crossing angle or changing $\beta^*$ and bunch intensity to get weaker LRIs. Here we study how the DA changes with the number and transverse separation of LRIs and explore scaling laws. We adjusted the bunch spacing and crossing angle to vary the number of LRIs over the range 12 to 48 at different constant separations from $9\sigma$ to $17\sigma$. The left plot in Fig. 12 shows the averaged DA as a function of the number of LRIs for three different separations. At each separation, the decrease in DA with increasing number $N_{LR}$ of LRIs can be fitted by a power law of the form shown in the following formula:

$$A_{dyn} = A + B(N_{LR} + C)^{-p}, \qquad (9)$$

where the parameters ($A$, $B$, $C$, $p$) depend in principle on the separation and $A_{dyn}$ is the averaged dynamic aperture. The fits show that parameter $A$ increases linearly with the separation but remarkably the other parameters are nearly independent of the separation. Specifically we find that the power law exponent $p=4.8$ is the same for all separations. The other parameters for three cases have the values: $A$=(6.9, 9.7, 14.4), $B$=(1.0, 1.0, 1.0) $\times$ $10^8$, $C$=(19.3, 20, 21.6). The three numbers in each bracket correspond to the LRI separation of $9\sigma$, $12\sigma$ and $17\sigma$, respectively. The fit in all three cases has a correlation coefficient is in the range $R=0.88 - 0.95$, where $R=1.00$ denotes a perfect fitting. The fact that only the parameter $A$ has a clear dependence on the separation suggests that Eq. (9) can be used to quickly estimate the DA for any other distribution of separations and number of interactions. The left plot in Fig. 12 also shows that the DA at a given separation does not change much for $N_{LR}$ sufficiently large. This observation suggests that an effective strategy to have a larger DA would be to have a greater number of LRIs at larger separations rather than fewer LRIs at smaller separations.

The right plot in Fig. 12 shows the DA variation with the transverse separation $r_{sep}$ for four values of $N_{LR} =(12, 24, 36, 48)$. In each case, the DA increases linearly with the transverse separation shown in the following formula:

$$A_{dyn} = D\, r_{sep} + E\,.\qquad(10)$$

The fits show that for the last three $N_{LR}$ values, the slope $D$ has the same value while the intercept $E$ decreases with increasing $N_{LR}$. The specific values for the four separations are $D=(0.60, 0.91, 0,91, 0.91)$ and $E=(8.26, 0.53, -0.68, -1.36)$. Here the correlation coefficient $R$ is in the range $0.94 – 0.99$. The small separation between the parallel lines for all $N_{LR} \geq 24$ reinforces the point above that the DA at a fixed separation does not change much with increasing $N_{LR}$ for $N_{LR}$ large enough.

As an example of the use of these laws, the DA of the entire ring could be found by using a weighted average of the DA at a discrete number of separations. This in turn could be used for quick estimates of the DA in other scenarios with a different distribution of transverse separations without a full tracking simulation for each case. These scaling laws also show that increasing the separations is more effective in increasing the DA rather than reducing the number of long-range interactions and this will be considered in more detail in Section V.

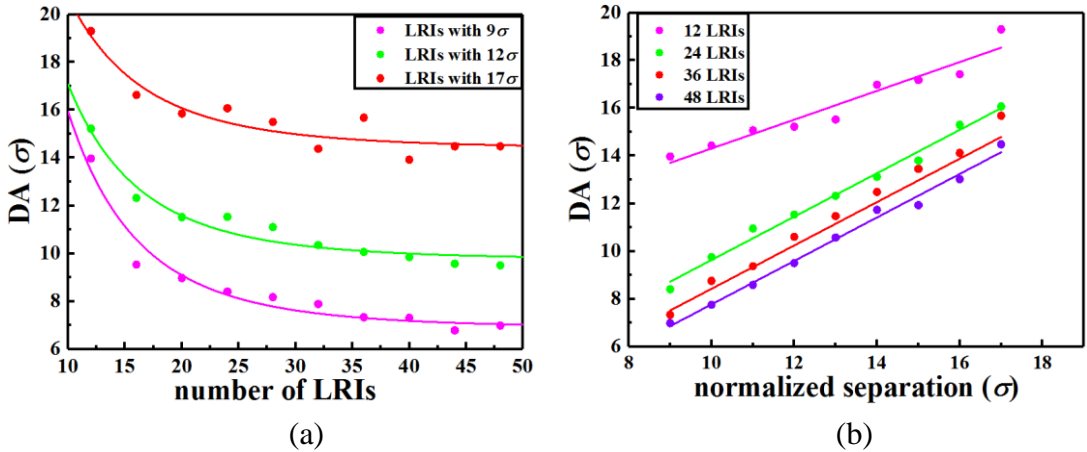

(a)          (b)

FIG. 12. Averaged DA plots with different numbers of LRIs (a) and separations (b)

## IV. TUNE SCAN ANALYSIS

To effectively operate accelerators, working points should be carefully picked otherwise the beam may be sensitive to errors and easily lost. Initially, the fractional tunes of SPPC are the same as in the LHC design, but now we do a tune scan to find some good working points. We investigated the DA with different fractional tunes and kept the integer part as before using the code BBSIM. In the tune space, the resonance free spaces are wider along the diagonal, thus the tune was scanned along lines parallel to the diagonal from (0.10, 0.10) to (0.46, 0.46) with a step size of 0.01. The different tune splits ($\pm 0.01$, $\pm 0.02$) were chosen to study how transverse coupling affects the dynamic aperture. The initial longitudinal distribution in $z$ was a Gaussian truncated at $2\sigma_z$. The chromaticity was still corrected to 1.00 here.

For a tune split of 0.01, Fig. 13 shows the averaged DA versus the horizontal fractional tune without and with nominal full crossing angle of 110 μrad. We notice that independent of the crossing angle and momentum deviation, the DA declines rapidly when the fractional tunes approach the low order resonances at 0.2, 0.25, and 0.33 while the crossing angle causes a steep drop additionally at 0.4. At three of the four tunes the DA with crossing angle and momentum deviation drops to less than $2\sigma$. This shows that the BB interactions drive the 3rd, 4th and 5th order resonances strongly. We also notice that with the crossing angle and zero initial momentum deviation, the DA is generally smaller than without the crossing angle. This is because the synchrotron oscillations and synchro-betatron resonances driven by the crossing angle cause this reduction and the drop increases with increasing values of the initial momentum deviation [22].

In addition, when there is a crossing angle, the beam-beam parameter decreases from 0.015 to 0.013, then the tune footprint will be closer to the dangerous sum resonances and this is another factor limiting beam stability.

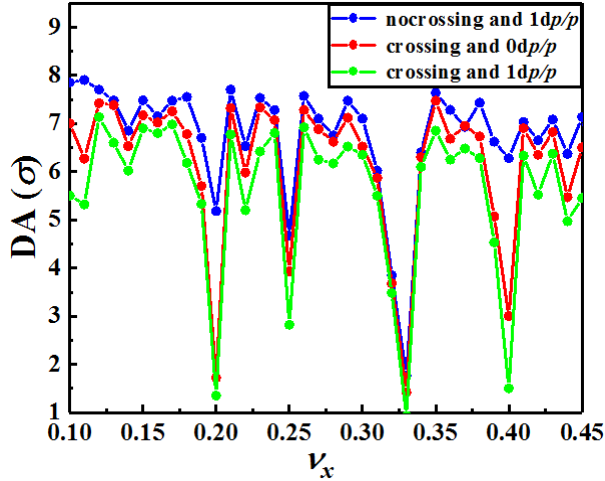

FIG. 13. Averaged DA with different $v_x$ nearing the coupling resonance with $v_y = v_x + 0.01$. Red: with a crossing angle and no d$p/p$ (momentum deviation); green: with crossing angle and d$p/p = \sigma_p$ (rms momentum spread); blue: without crossing angle and d$p/p = \sigma_p$.

A larger tune split 0.02 was also tried to avoid transverse coupling and increased the tune distance with respect to the nearby sum resonances. For a tune split of 0.02, the DA increases except near 3rd order resonance. Thus the 3rd order resonances play the dominant role in limiting beam stability compared with 4th and 5th order resonances. Finally, the tune scan reveals there are 3 tunes where the DA increases by $1.5\sigma$ to $7\sigma$ and 3 others where the increase in DA is about $1\sigma$. In general the smallest dynamic aperture is about $1\sigma$ smaller than the averaged DA. Table V shows the DA of the 6 best tunes and the nominal tune. The tunes identified here could serve as good initial candidates for a working point to be examined with more detailed studies.

TABLE V. Dynamic aperture of 6 best tunes and the nominal tune.

| Tunes | Averaged DA ($\sigma$) | Smallest DA ($\sigma$) |
|---|---|---|
| (0.12,0.13) | 7.13 | 6.25 |
| (0.17,0.19) | 7.12 | 6.25 |
| (0.27,0.26) | 7.02 | 6.00 |
| (0.37,0.35) | 6.70 | 5.75 |
| (0.19,0.17) | 6.57 | 5.75 |
| (0.38,0.37) | 6.50 | 6.25 |
| (0.31,0.32) | 5.50 | 4.75 |

In order to examine the reasons for the better tunes, we studied the FMA plots. Figure 14 shows the FMA plots in amplitude and frequency space for the nominal tune and two better tunes. We observe from the amplitude space plots that the tune diffusion is smaller at the better tunes for particles amplitude from $6\sigma$ to $8\sigma$. The plots in the tune space show that the footprint for the nominal tune is crossed by the 3rd order resonances, the footprints for the other two tunes are crossed by 4th and 6th order sum resonances respectively. As expected, lower the order of the important resonance, smaller is the dynamic aperture.

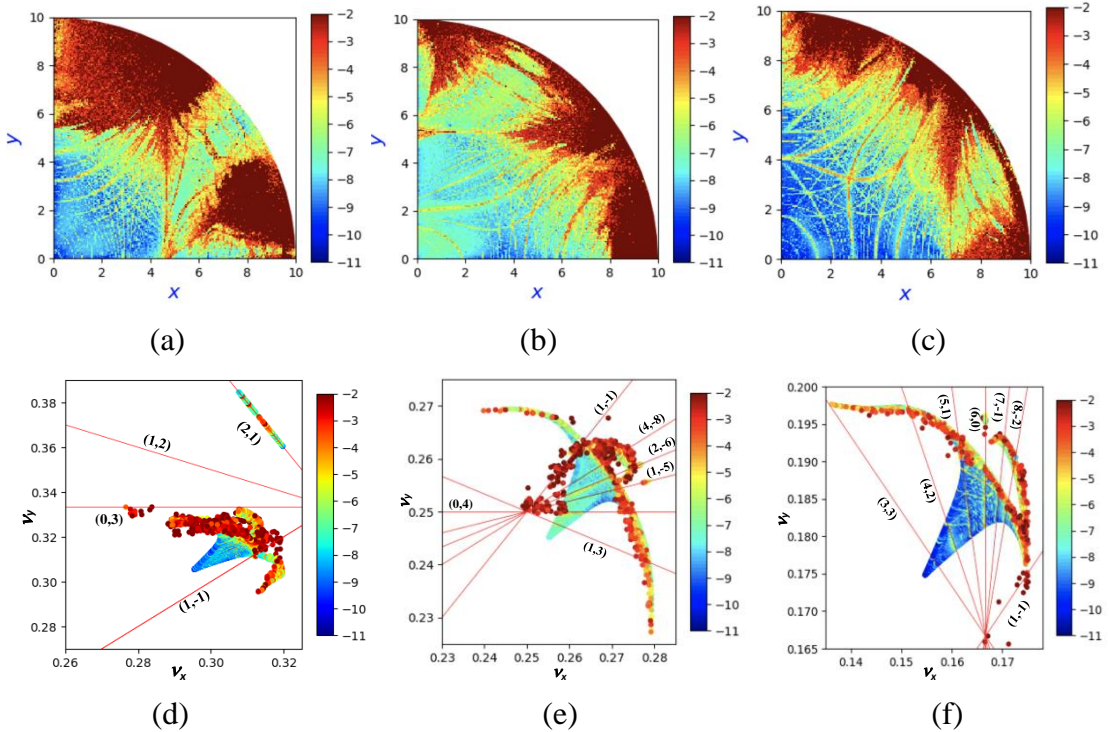

FIG. 14. FMA plots in the amplitude (top) and frequency (bottom) space for the nominal tune (left) and a better tune (middle) of (0.27, 0.26) and another better tune

(right) of (0.17,0.19). The numbers in brackets (*m, n*) define the resonances $m\nu_x+n\nu_y=p$, where *p* is an integer.

## V. LONG-RANGE INTERACTION MITIGATION WITH MODIFIED BETA STAR AND CROSSING ANGLE

While the above tune scan shows that better tunes than the nominal tunes could be found, the increase in DA is not significant, so other ways of improvement need to be found. One option is to increase the BB transverse separation. We recall that the BB kick depends on the transverse separation normalised by the rms transverse size of the strong beam. There are two ways to control the BB transverse separation. One is to adjust the crossing angle that can change the absolute transverse separation and another one is to adjust *β\** that can change the beam size.

In the baseline design, the BB transverse separation is $12\sigma$ at the first parasitic interaction with the full crossing angle of 110 μrad and *β\** of 0.75 m. Here we increased the separation at the first parasitic interaction from $12\sigma$ to $20\sigma$ by adjusting the crossing angle and *β\** together. The separation at the other LRIs would also increase synchronously but not necessarily by the same amount. As in Section II, we considered two other values of *β\**=0.5 m and 1.0 m. The working point was chosen to be at the nominal tunes while the chromaticity in all cases was corrected to 1.0 and the momentum deviation was one time the rms momentum spread.

Figure 15 shows the averaged DA versus the BB separation at the first parasitic interaction for the three values of *β\**. It is expected that with the increasing separation, the averaged DA increases for all values of *β\**. The desired goal is to achieve a DA≥$12\sigma$, considering that important errors such as triplet nonlinearities, orbit errors etc are not included. For each *β\**, there is a $6\sigma$ DA improvement when the BB separation increases from $12\sigma$ to $20\sigma$, which is the separation required for the DA goal of at least $12\sigma$. Since the crossing angle scales as $1/\sqrt{\beta^*}$, to reach the DA goal the crossing angle increases from 160 μrad at *β\**=1 m to 184 μrad at *β\**=0.75 m and to 221 μrad at *β\**=0.5 m. We find that the DA is independent of the *β\** only if the scaled transverse separation stays the same. This may change when the nonlinear fields in the IR quadrupoles are included.

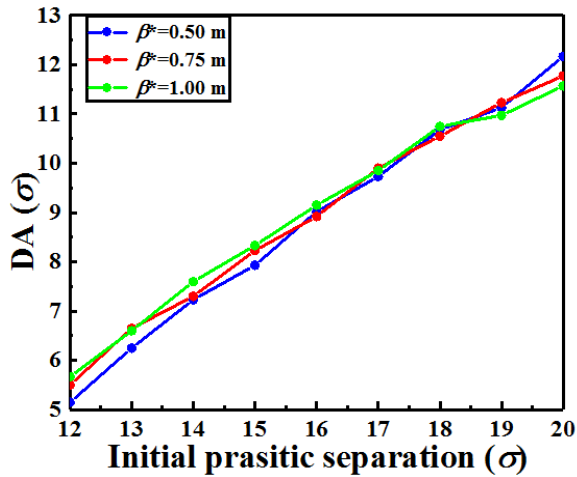

FIG. 15. The averaged DA for different $\beta^*$ values with increasing 1st parasitic separation that is normalised by its horizontal beam size. The momentum deviation is one time the rms momentum spread for all those cases.

Although the lager crossing angle brings better beam stability, it has many negative effects, e.g., lower luminosity, lower relative physical aperture for the inner quadrupole triplets, enhanced synchro-betatron resonances and worse nonlinearities for the inner quadrupole triplets. Especially for $\beta^*=0.5$ m, it was shown in Table III that half the luminosity is lost. At this time the plan is to use crab cavities to recover the luminosity and minimize synchro-betatron resonances, assuming their successful operation in the LHC. However, the smaller relative physical aperture and worse nonlinearities of the triplets will remain important limitations. It looks likely that the active compensation of the LRIs will need to be considered to improve beam stability, almost certainly for $\beta^*\leq 0.5$ m to allow operation at smaller crossing angles and perhaps also for larger $\beta^*$ values.

## VI. CONCLUSIONS

The SPPC baseline design with a crossing scheme that also corrects the dispersion at the IP has been described and proved to be robust against orbit errors. We studied the optics with different $\beta^*$ and find that $\beta^*\leq 0.5$ m is excluded in the present geometry by optics constraints, such as the available physical aperture. We carried out weak-strong beam-beam simulations for the SPPC and studied the beam performance from footprint and FMA plots and dynamic aperture calculations. It is demonstrated that the long-range interactions are the main factor limiting particle stability. Amongst all of these interactions, the ones with the smallest separations are particularly destructive. Scaling laws are obtained showing the dependence of the dynamic aperture on the separation and on the number of interactions. These show that changing the geometry to increase the transverse separations while allowing more interactions if necessary would be helpful. A tune scan analysis was done to find better tunes and it is noticed that the 3rd, 4th and 5th order resonances are driven with the BB interactions. A more useful option to increase the dynamic aperture significantly by adjusting the crossing angle and β* together to increase BB transverse separation is shown to be successful. However, the negative consequences from a larger crossing angle will require the active compensation of the long-range interactions and possibly crab cavities to recover the luminosity. We also mention that the studies reported here need to be repeated with a complete description of errors and the inclusion of radiation damping effects. These and other topics will be subjects of future studies.

## ACKNOWLEDGMENT
We would like to thank all colleagues in the SPPC collaboration. We specially thank Prof. Philippe Piot and Northern Illinois University for their hospitality in hosting the visit of the first author L.W, and China Scholarship Council for supporting her to study at NIU and Fermilab. Fermilab is operated by the Fermi Research Alliance, LLC under U.S. Department of Energy contract No. DE-AC02-07CH11359. The work of L.W and J.T was supported by the National Natural Science Foundation of China (Projects No. 11575214 and No. 11527811).